\documentclass[useAMS,usenatbib]{mn2e}

\usepackage{graphics}

\title{Luminosity function of contact binaries based on
the ASAS survey}
\author[Slavek M. Rucinski]
{Slavek M. Rucinski$^{1}$\thanks{e-mail: rucinski@astro.utoronto.ca}\\
$^{1}$Department of Astronomy and Astrophysics, University of Toronto\\
David Dunlap Observatory, 
P.O.Box 360, Richmond Hill, Ontario, Canada L4C~4Y6}

\date{Accepted --.
      Received -- ;
      in original form --}

\pubyear{2006}

\begin{document}

\maketitle

\label{firstpage}

\begin{abstract}
The luminosity function for contact
binary stars of the W~UMa-type is evaluated on the basis of the 
ASAS photometric project covering all stars south of
$\delta= +28^\circ$ within a magnitude range $8 < V < 13$. 
Lack of colour indices enforced a limitation to 
3373 systems with $P<0.562$ days (i.e.\ 73\% of all 
systems with $P<1$ day) where a simplified $M_V (\log P)$ 
calibration could be used. 
The spatial density relative to the main sequence FGK stars
of 0.2\%, as established previously from the Hipparcos 
sample to $V=7.5$, is confirmed. 
While the numbers of contact binaries in the ASAS survey 
are large and thus the statistical uncertainties small, 
derivation of the luminosity function required a 
correction for missed systems with small amplitudes 
and with orbital periods longer than 0.562 days; 
the correction, by a factor
of $3 \times$, carries an uncertainty of about 30\%. 
\end{abstract}


\section{Introduction}  
\label{sect:intro}

ASAS -- All Sky Automated Survey\footnote{See:
http://www.astrouw.edu.pl/\~{}gp/asas/asas.html and 
http://archive.princeton.edu/\~{}asas/ }, is a long 
term project dedicated to detection and monitoring 
variability of bright stars using small telescopes. 
Recently \citet{BP2006} presented results of the 
$V$-filter observations obtained at the Las Campanas Observatory 
with a single 7~cm telescope. 
More than 50 thousand variables brighter 
than 14 magnitude were discovered in about three fourth of
the sky with $\delta < +28^\circ$, 
among them 5,384 contact binary stars. 

This paper addresses the matter of the spatial density
of contact binaries of the W~UMa type ($P<1$ day, 
from now on called ``EW'') as indicated by the ASAS survey 
data within its magnitude limits of $8 <V< 13$. The relative
spatial density (in relation to the main sequence FGK
dwarfs) is evaluated via the luminosity function 
for the absolute magnitude range $1.5 < M_V < 5.5$. 

We are very far from having a volume limited sample of 
EW binaries. In fact, the main problem is the completeness
to a given magnitude limit. At present, the only sample
which appears to include all EW binaries 
with photometric variability amplitudes $>0.05$ mag.\ 
is the shallow Hipparcos sample to $V=7.5$ 
\citep{Rci2002} (from now on called the ``Hipparcos
sample'' or the ``7.5 mag.\ sample'') consisting of
35 objects, with 1/3 in that number being new detections.
This sample suggests the relative spatial density of 
the EW binaries at a level of 0.2\% of the FGK
main sequence stars, but this fraction requires 
a confirmation by a larger sample,
particularly for $M_V>+3.5$ where the Hipparcos
sample contained very few objects.

\section{Amplitude distribution and missing systems}
\label{sect:ampl}

Spatial density estimates of variable stars 
are crucially dependent on the
detection limitations: Photometric variability must be
some 5 times larger than the photometric error, $\sigma$,
for detection while another margin of 
$>10 \, \sigma$ is needed for proper characterisation and 
assurance that the star is not one of rather common 
$\delta$~Sct or $\gamma$~Dor short-period pulsating 
main-sequence stars \citep{Rci2004}.
Theoretical predictions \citep{Rci2001} indicate
that low-amplitude contact binaries should be very common
as the amplitude distribution is expected to rise for
amplitudes going to zero. 

\begin{figure}
\begin{center}
\rotatebox{90}{\scalebox{0.37}{\includegraphics{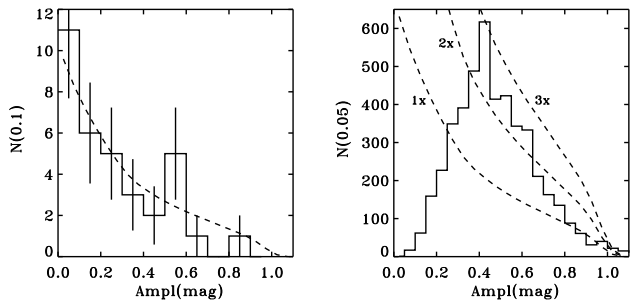}}}
\caption{\label{fig1}
The amplitude distributions for 35 contact binaries of
the Hipparcos 7.5 mag.\ survey (the left panel) and for  
4640 binaries of the ASAS survey (the right panel) 
with $P<1$ day are shown together with the expected 
distribution for randomly distributed orbital inclinations
and a flat distribution of mass ratios (the dashed line).
In both panels the theoretical curve is normalized to
the total number of systems, although we show it also scaled
up by $2 \times$ and $3 \times$ in the right panel.
}
\end{center}
\end{figure} 

EW binaries discovered in the ASAS survey show a 
different amplitude distribution from that of the
Hipparcos sample. Figure~\ref{fig1} gives the distribution
for the Hipparcos sample (left panel) and for the ASAS
sample (right panel), both 
compared with the theoretical amplitude distribution
expected for random orbital inclinations and for a flat
mass-ratio distribution, $Q(q) = {\rm const}$
\citep{Rci2001}. The latter assumption is made for 
lack of any evidence to the contrary and needs verification.
Also, both observational samples include a small
admixture of contact binaries with unequally deep 
eclipses (sometimes called EB); the presence of these binaries 
tends to elevate the high amplitude end of the distribution,
so that the agreement of the Hipparcos and the theoretical
distributions may be fortuitous. 

It is easily noticeable in Figure~\ref{fig1} that while
the Hipparcos sample follows the theoretical prediction, but
suffers from the low number statistics., 
the ASAS distribution lacks systems with amplitudes $<0.4$ mag. 
This can be understood in terms of the detection 
selection effect with a relatively high detection threshold. 
For a direct comparison of the Hipparcos and ASAS samples, 
the numbers estimated from the latter must be 
multiplied by a correction factor of about $2 - 2.5 \times$ 
(see Figure~\ref{fig1}), with 
a considerable uncertainty of some 25\% or so.

Compared with the theoretical and Hipparcos distributions, 
the ASAS amplitude distribution appears to have some excess
of systems with amplitudes in the range of $0.4 < a < 0.6$
and too few systems with amplitudes $>0.6$ mag. 
The absence of large amplitudes may be in the 
photometric blending of images and 
thus a ``dilution'' of the 
amplitudes because of the large images given by the 7~cm
telescope, particularly when coupled with the now 
recognized prevalence ($>50$\%) of triple and multiple 
systems with contact binaries \citep{PR2006}.

\section{Absolute magnitude calibration}
\label{sect:abs-mag}

The previously proposed absolute magnitude calibration
for contact binaries \citep{Rci1994,RD1997,Rci2004},
is a linear one:
$M_V = a_P \log P + a_{BV} (B-V)_0 + a_0$, where
$(B-V)_0$ is the reddening corrected colour index
(which can be replaced by say $(V-I)_0$); it can
predict $M_V$ to about 0.2 -- 0.25 mag.
It has been used successfully for many applications,
particularly for confirming or disproving membership
of contact binaries in stellar clusters
\citep{Rci1998,Rci2000}. The main basis for the
calibration is the Hipparcos sample \citep{Rci1994,RD1997}.
This sample has been carefully scrutinized for multiplicity 
and for small, but measurable amounts of
the local interstellar (IS) reddening. The former requirement 
is related to the high frequency of triple systems 
among contact binaries while the latter
stems from the availability of sensitive indicators of
small amounts of local reddening such as the linear
polarization, EUV extinction or IS cirrus emission.

\begin{table}
\begin{scriptsize}
\caption{\label{tab1}
The sample used to establish $M_V=M_V(\log P$ calibration.
VW~Cep is the only system where corrections for the third
component with $L_3/L_{12}=0.08$ has been applied. 
The last four systems were not used, see the text.
}
\begin{center}
\begin{tabular}{rrrrrrr} 
Name & $V_{max}$ & $B-V$ & $E_{B-V}$ & $P (d)$ & $M_V$ & $\varepsilon M_V$ \\
  VW Cep &   7.38 &   0.82 &   0.00 &   0.2783 &   5.17 &   0.06 \\
  OU Ser &   8.25 &   0.62 &   0.01 &   0.2968 &   4.41 &   0.13 \\
  SX Crv &   8.95 &   0.52 &   0.04 &   0.3166 &   4.01 &   0.24 \\
  YY Eri &   8.16 &   0.66 &   0.00 &   0.3215 &   4.43 &   0.14 \\
   W UMa &   7.76 &   0.62 &   0.00 &   0.3336 &   4.28 &   0.11 \\
  GM Dra &   8.65 &   0.48 &   0.03 &   0.3387 &   3.59 &   0.19 \\
V757 Cen &   8.40 &   0.65 &   0.02 &   0.3432 &   4.10 &   0.17 \\
V781 Tau &   8.55 &   0.58 &   0.02 &   0.3449 &   3.94 &   0.24 \\
  GR Vir &   7.81 &   0.56 &   0.01 &   0.3470 &   4.15 &   0.14 \\
  AE Phe &   7.56 &   0.64 &   0.00 &   0.3624 &   4.12 &   0.09 \\
  YY CrB &   8.64 &   0.62 &   0.01 &   0.3766 &   3.89 &   0.16 \\
V759 Cen &   7.47 &   0.53 &   0.02 &   0.3940 &   3.41 &   0.13 \\
  EX Leo &   8.17 &   0.53 &   0.00 &   0.4086 &   3.13 &   0.24 \\
V566 Oph &   7.47 &   0.41 &   0.03 &   0.4096 &   3.10 &   0.17 \\
  AW UMa &   6.84 &   0.33 &   0.01 &   0.4387 &   2.71 &   0.13 \\
  CN Hyi &   6.57 &   0.41 &   0.01 &   0.4561 &   2.72 &   0.08 \\
  TY Men &   8.11 &   0.29 &   0.01 &   0.4617 &   1.94 &   0.23 \\
  RR Cen &   7.32 &   0.34 &   0.03 &   0.6057 &   2.17 &   0.19 \\
  IS CMa &   6.87 &   0.31 &   0.01 &   0.6170 &   1.84 &   0.15 \\
V535 Ara &   7.15 &   0.30 &   0.02 &   0.6293 &   1.83 &   0.22 \\
   S Ant &   6.29 &   0.30 &   0.01 &   0.6484 &   1.88 &   0.12 \\
\end{tabular}
\end{center}
\end{scriptsize}
\end{table}

\begin{figure}
\begin{center}
\scalebox{0.5}{\includegraphics{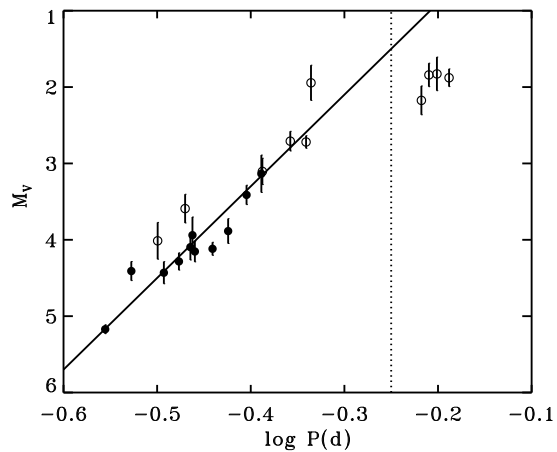}}
\caption{\label{fig2}
The period -- luminosity calibration for the 21 EW systems
with good Hipparcos parallaxes and free of triple-system
complications. The open and closed circles mark systems
bluer and redder than $(B-V)_0=0.5$, respectively. The line
is given by: $M_V=-1.5-12 \log P$ with the orbital period
$P$ in days. It was derived from 17 systems with $\log P < -0.25$.
}
\end{center}
\end{figure} 

The calibrating sample is given in Table~\ref{tab1}.
It consists of 21 EW from the 
Hipparcos 7.5 mag.\ sample \citep{Rci2002}. Except for
for the orbital periods (known practically without any error)
and for the parallax and parallax error data taken from the
Hipparcos mission catalogue (not listed here,
but available in previous tabulations), the remaining
data have been modified: The known triple systems
have been removed (with the exception of VW~Cep, 
which is important for low $M_V$; its 
ratio $L_3/L_{12}$ is well known) and 
the magnitudes and colour indices were corrected
for newly derived values of the IS reddening. The 
reddening is given to 0.01 mag.\ for consistency 
with typical accuracy of $B-V$, but its value is 
frequently known much better.
The table should be regarded as an improvement
over the previous similar tabulations 
\citep{RD1997,Rci2002}
in the treatment of the IS reddening and in the careful
screening for triple components.

The ASAS data presented by \citet{BP2006} are in the $V$ band. The
colour indices are currently unknown, but will be available 
in the future. We derive here a simplified
$M_V=M_V(\log P)$ calibration utilizing the known loose
correlation of the periods and colour indices, as noticed
long time ago by \citet{Egg1967}.
We claim that a very simple relation of the form
$M_V = a_0 + a_P \log P$ 
may suffice in situations like the current one when
an additional limit is added on the length of the
orbital period. The very steep dependence is
indeed visible in Figure~\ref{fig2}. Some scatter is due to
a range in the reddening corrected colour index, 
$0.28 < (B-V)_0 < 0.82$; the spread is noticeable in the 
figure where different symbols are used for $(B-V)_0$ 
above and below 0.5. Four binaries with $\log P > -0.25$
are all blue, so a decision was made to use only 17 binaries
with periods shorter than this limit.
By using the bootstrap sampling technique of repeated 
least-squares solutions, we derived a simplified relation: 
$M_V = -1.5 (\pm 0.8) -12.0 (\pm 2.0) \log P$, with
$\sigma = 0.29$ for a single object. The errors
of the coefficients have been estimated from the 
bootstrap sampling distributions encompassing 67\% of the results.

If a simplified calibration works, a question arises whether
the colour term is needed at all. The data have well 
established errors $\varepsilon M_V$ derived from the parallax errors,
so one can check if $\chi^2 = \sum{(O-C)^2/\sigma^2}$ is approximately
equal to the number of degrees of freedom. 
The 3-term linear fit $M_V=a_0 + a_P \log P + a_{BV} (B-V)$ 
gives $\chi_3^2 = 13.4$ for $\nu= m-3=14$ degrees of
freedom, thus indicating a basically perfect fit and
the need of a 3-term fit\footnote{The coefficients for this
sample are: 
$a_P=-5.8 \pm 1.1$, $a_{BV}=+2.75 \pm 0.45$, $a_0=-0.31 \pm 0.32$.
Because of the period -- colour ($a_P - a_{BV}$) correlation,
the $M_V$ predictions are practically the same
as for coefficients in \citet{Rci2000,Rci2004}. For
consistency with the previous results,
it is recommended to continue using the old coefficients.}.
The simplified 2-term linear fit, 
$M_V = -1.5 -12.0 \log P$, 
gives $\chi_2^2 = 52.9$ for $\nu = m-2=15$ degrees of freedom,
confirming that indeed three terms are needed. However,
we can invert the problem and ask if the {\it decrease\/} 
in $\chi^2$ due to the addition of the colour term is significant.
It appears that the difference between $\chi_2^2$ 
and $\chi_3^2$, tested with the $F$ Snedeckor distribution
\citep{Ham1964} suggests a probability of an {\it accidental\/} 
improvement at the level of 37\%, quite a distance from the normally
desired $<5$\%, so it would appear that the third term is
not needed after all. Thus, the situation is somewhat confused.
However, a consideration of the involved physics
strongly indicates that a colour term should be
present in the calibration unless it is not available,
as in the current application.

\begin{figure}
\begin{center}
\scalebox{0.5}{\includegraphics{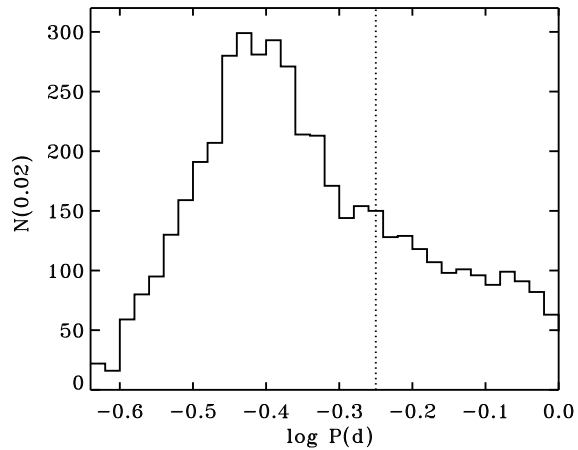}}
\caption{\label{fig3}
The period distribution for the ASAS sample of 4640 
EW systems with $P<1$ day. The $M_V$ calibration 
derived here applies to $\log P < -0.25$
and this limit is marked by the vertical dotted line. 
}
\end{center}
\end{figure} 

\section{Distances and spatial density}
\label{sect:dens}

The limitation to periods $\log P < -0.25$ or $P < 0.562$
days means that from now on we will consider only the EW
binaries from the main peak of period distribution,
as shown in Figure~\ref{fig3}.
Thus, out of the full sample of 5384 EW binaries in the ASAS survey, 
from now on
we will discuss the results for 3377 systems or 63\% of the whole 
sample or 73\% of the sub-sample of 4640 systems with $P<1$ day. 
For a direct comparison of numbers with the Hipparcos
sample, a correction factor of 1/0.73=1.37 must be applied. 
This factor, combined with the under-counting due to loss of
low amplitude binaries (Section~\ref{sect:ampl}) implies
that the ASAS counts must be multiplied by a factor of about
$3 \times$ carrying an uncertainty of about 30\%.

\begin{figure}
\begin{center}
\scalebox{0.5}{\includegraphics{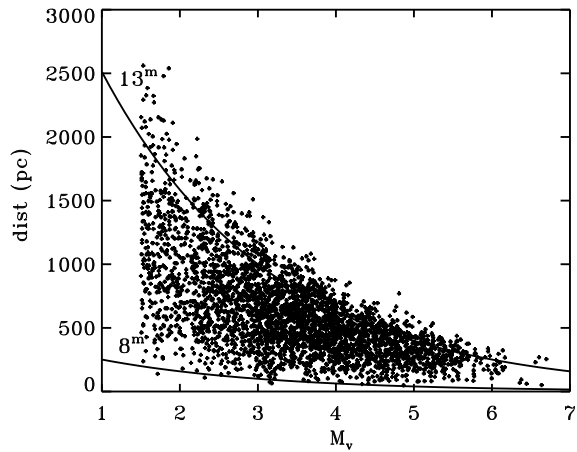}}
\caption{\label{fig4}
Distances to 3377 EW systems with $\log P < -0.25$ plotted 
versus $M_V$. The distance limits for $V=8$ and $V=13$ are
shown by lines.  
}
\end{center}
\end{figure}

With the known values of $M_V$, the distances are simply 
calculated from $d={\rm dex}((V-M_V+5)/5)$, 
where we neglect the interstellar
reddening. As we can see in Figure~\ref{fig4}, the distances are 
moderate, so this assumption is acceptable except
for the bins of higher luminosities where the distances extend beyond
a few hundred parsecs. Thus, we expect that for $M_V < +3.5$
our density estimates are too low. However, we are
interested mostly in the luminosity function for fainter contact
binaries where the Hipparcos sample \citep{Rci2002} ran out
of low luminosity objects.

\begin{table}
\begin{scriptsize}
\caption{\label{tab2}
Table 2:
The luminosity function for ASAS sample
of EW systems with $\log P <-0.25$. 
$n$ is the number of EW systems in 
a 1-mag.\ wide bins centred at $M_V$ within the distances
$r_1$ and $r_2$. $r_1$ corresponds to $V=8$ and $M_V-0.5$
while $r_2$ corresponds to $V=13$ and $M_V+0.5$. $\phi$
is the luminosity function in units of $10^{-7}$ stars/pc$^3$; 
its error $\varepsilon \phi$ is evaluated from $\sqrt{n}$.
}
\begin{center}
\begin{tabular}{rrrrrr} 
$M_V$ & n \ \ & $r_1$ (pc)& $r_2$ (pc)& $\phi$ \ \ & $\varepsilon \phi$  \ \ \\
2.0 & 489 & 199.5 & 1258.9 &  0.80& 0.04\\
3.0 & 729 & 125.9 &  794.3 &  4.74& 0.18\\
4.0 & 564 & 79.4  &  501.2 & 14.61& 0.62\\
5.0 & 229 & 50.1  &  316.2 & 23.62& 1.56\\
6.0 &  26 & 31.6  &  199.5 & 10.67& 2.09\\
\end{tabular}
\end{center}
\end{scriptsize}
\end{table}

\begin{figure}
\begin{center}
\scalebox{0.5}{\includegraphics{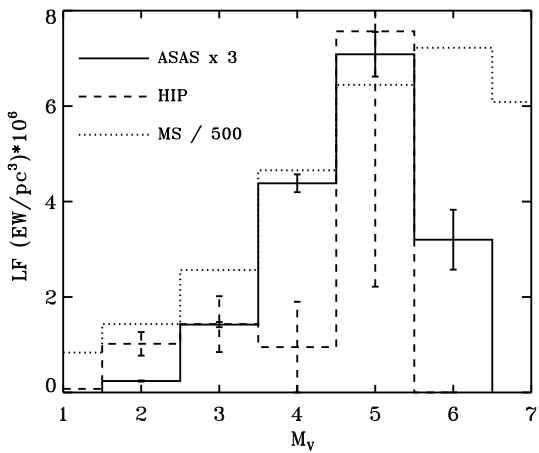}}
\caption{\label{fig5}
The luminosity function for the EW systems with $\log P <-0.25$
in the ASAS survey (the continuous line) is compared with
the results based on the Hipparcos sample extending to
$V_{max} < 7.5$ (the broken line) and with the solar 
neighbourhood data scaled down by factor $500 \times$ (the
dotted line). Note that for the direct comparison with the
Hipparcos results, the ASAS counts have been multiplied by a
factor of $3 \times$. The error bars show the mean standard 
errors resulting from the statistical uncertainties only
and do not include the systematic uncertainty of the 
ASAS detection correction factor estimated at about 30\%.
}
\end{center}
\end{figure}

The ASAS photometric survey covered the 5-magnitude interval $8 < V < 13$. 
The two magnitude limits, expressed as distances, are shown as lines in
Figure~\ref{fig4}. The other limitation comes from our $M_V$
calibration (Section~\ref{sect:abs-mag})
which is valid within $1.5 < M_V < 5.5$. 
For each 1-mag.-wide bin in $M_V$, 
we found the near and far limits 
($r_1$ and $r_2$ in Table~\ref{tab2}),
and simply counted the systems in the respective
volume $4 \pi/3 \times C \times (r_2^3-r_1^3)$ assuming that
the survey South of $\delta = +28^\circ$
covered $C=0.735$  of the whole celestial sphere.
The results are given in Table~\ref{tab2}, while the luminosity 
function is shown in Figure~\ref{fig5}.

\section{The luminosity function}

Figure~\ref{fig5} compares the ASAS luminosity function,
after correction by $3 \times$ 
(for systems missed because of the amplitude detection 
selection and of the removal of systems with $P > 0.562$ days), 
with the Hipparcos luminosity function and
with the Main Sequence luminosity function \citep{Wie1983} 
scaled down by a factor of 500 times. The ASAS and
Hipparcos sample results are complementary, 
with the ASAS sample improving the
luminosity function within the interval $3.5 < M_V < 5.5$
where the Hipparcos sample had very few objects.
While the luminosity function is better 
defined in the statistical sense 
with hundreds of binaries per $\Delta M_V$ bin, the
correction for the systematic effects 
by a factor of $(3 \pm 1) \times$ leaves a rather 
wide margin for improvement.

The comparison of the luminosity functions in
Figure~\ref{fig5} confirms the 
previously derived relative spatial frequency of contact 
binaries in the solar neighbourhood of 0.2\%. Although  
the relative frequency of bright contact binaries,
$M_V < +3.5$, appears low at about 0.1\%, the lower
numbers of intrinsically bright binaries
may be caused by our neglect of the IS reddening and
extinction and by the inapplicability of the simplified
$M_V$ calibration. The relative frequency reaches 
0.2\% for the interval $+3.5 < M_V < +5.5$, but
then drops to 0.1\% in the $M_V = +6$ bin,
and to zero for still fainter objects. The sharp
decline in the numbers of faint systems is related
to the well established period cut-off at 0.215 -- 0.22 
days\footnote{The system CC~Com delineated the cut-off at
0.2207 d.\ for a long time, but a new discovery
in 47~Tuc by \citet{Wel2004} has moved it to a still
shorter period of 0.2155 d. Short periods are expected
and indeed observed in globular clusters \citep{Rci2000}.}
which still remains unexplained. It should be
noted that the relative frequency may be higher
in the central galactic disk at about 1/130 \citep{Rci1998}
although this particular estimate suffered from
photometric blending effects in the direction of the
Galactic Bulge which may have raised the frequency
by a factor of about $2 \times$.

The author would like to express his thanks 
to the team of B.\ Paczy\'{n}ski, D.\ Szczygie{\l}, 
B.\ Pilecki and G.\ Pojma\'{n}ski for
enthusiastic and rapid sharing of their data
and to Grzegorz Pojma\'{n}ski for making ASAS such an 
useful variable-star study tool.

Support from the Natural Sciences and Engineering 
Council of Canada is acknowledged with gratitude.

\end{document}